\documentclass[prl,twocolumn,floatfix,nobibnotes,superscriptaddress,prb]{revtex4}
\usepackage{amsmath}
\usepackage{graphicx}

\begin{document}
\title{Entropy driven stabilization of energetically unstable crystal structures explained from first principles theory}
\author{P. Souvatzis}
\affiliation{Department of Physics, Uppsala University,
Box 530, SE-75121, Uppsala, Sweden}
\author{O. Eriksson}
\affiliation{Department of Physics, Uppsala University,
Box 530, SE-75121, Uppsala, Sweden}
\author{M. I. Katsnelson}
\affiliation{Institute for Molecules and Materials, Radboud
University Nijmegen, NL-6525 ED, Nijmegen, The Netherlands}
\author{S. P. Rudin}
\affiliation{Theoretical Division, Los Alamos National Laboratory,
Los Alamos, New Mexico 87545, USA }
\date{\today }

\begin{abstract}
Conventional methods to calculate the thermodynamics of crystals
evaluate the harmonic phonon spectra and therefore do not work in
frequent and important situations where the crystal structure is
unstable in the harmonic approximation, such as the body-centered
cubic (bcc) crystal structure when it appears as a high-temperature phase of many
metals. A method for calculating temperature dependent phonon spectra self consistently from first principles has been 
developed to address this issue. The method combines concepts from Born's inter-atomic self-consistent phonon approach with 
first principles calculations of accurate inter-atomic forces in a super-cell. The method
has been tested on the high temperature bcc phase of Ti, Zr and Hf, as representative examples,
and is found to reproduce the observed high temperature phonon frequencies with good accuracy. 
\end{abstract}

\maketitle

Many elements, alloys, and compounds appear in crystal structures
which should not be energetically stable. The inter-atomic
interactions places these systems at energy saddle points on the
potential surface for atomic positions corresponding to the
lattice sites of these structures rather than minima for
statically stable structures. The body centered cubic (bcc)
structure prevails as the simplest and best known example.
Although a stable structure at low temperatures for several
elements in the Periodic Table, bcc becomes unstable in the harmonic
approximation~\cite{born1, KRIS,Ye} for the group IVB elements, for the
rare-earth elements, for the actinides, and for several
alkaline-earth elements. Nevertheless, at elevated temperatures
the bcc structure emerges as the stable crystal structure for all
these elements. Zener considered this enigma long ago and proposed a possible explanation:
the large vibrational entropy of the bcc crystal structure makes it thermodynamically 
favourable at finite temperatures~\cite{zener}. Also, Grimvall {\it et al} \cite{GRIM} 
pointed out the importance of electronic entropy in the stabilization of the bcc crystal structure  
of the group IVB elements Ti and Zr.

So far no satisfactory, quantitative explanation has been
presented for this situation. Density functional theory (DFT)
~\cite{KOHN1} forms the basis of contemporary microscopic
solid state theory and allows, in principle, to calculate
different properties of crystals completely {\it ab initio},
without any fitting parameters. In particular, phonon spectra in the harmonic approximation
can be efficiently evaluated in this way~\cite{baroni}. However, for
the bcc phases mentioned above the phonon spectra in the
harmonic approximation reveal imaginary phonon frequencies of e.g. Zr\cite{ho1,ho2}
 for some wave-vectors, which shows that the bcc phase is from a lattice dynamics point of view unstable (hence 
these elements are energetically unstable, and are referred to as dynamically unstable in the bcc phase). A
straightforward calculation using DFT molecular dynamics
(MD)~\cite{PARINELLO} should in principle be able to reproduce the
stability of the bcc phase for the above discussed elements, since
MD implicitly include temperature effects. However, MD suffers
from the fact that obtaining reliable free energies implies a 
computationally very demanding task, which in many cases
make these types of calculations
intractable. 

We propose here a solution to this problem, which builds on a
self-consistent {\it ab initio} lattice dynamics (SCAILD)
approach. In this paper we describe the essential aspects of our
method and apply it to the problem of stability of the bcc phase
for the group IVB elements. Although several aspects of our
proposed theory have not been considered before we note that it
conceptually has similarities with the self-consistent phonon
approach by Born~\cite{born}, and that several other self-consistent methods \cite{KOELER,HOOTEN} 
have been developed in the past. We will show that the SCAILD theory
gives phonon spectra of the bcc phase of Ti, Zr and Hf which are
in agreement with observations~\cite{HEIMING,PETRY,TRAMENAU}. Although we will in the rest of
this manuscript focus on the group IV elements, we point out here
that what we provide is a general scheme which can be used for any
element and compound.

Self consistent phonon calculations are a natural extension of the theory of the
harmonic lattice, and we initiate our methodological description by first presenting the most important features of this theory. The Hamiltonian
\begin{equation}\label{eq:harmH1}
\mathcal{H}_{h} = \sum_{\mathbf{R}}\frac{\mathbf{P}_{\mathbf{R}}^2}{2M}+\frac{1}{2}\sum_{\mathbf{R},\mathbf{R}'}\mathbf{U}_{\mathbf{R}}\bar{\bar{\Phi}}(\mathbf{R}-\mathbf{R}')\mathbf{U}_{\mathbf{R}'},
\end{equation}
describes a harmonic lattice where $\mathbf{R}$ are the
equilibrium lattice positions of the atoms,
$\mathbf{U}_{\mathbf{R}}$ the displacements of the atoms,
$\mathbf{P}_{\mathbf{R}}$ the momentum of the atoms, $M$ the
atomic mass and $\bar{\bar{\Phi}}$ the inter-atomic force constant
matrices (Here the vectors $\mathbf{R}$ refere to the positions of a Bravais lattice). Diagonalizing the dynamical matrix
\begin{equation}\label{eq:dynm}
\bar{\bar{\mathcal{D}}}(\mathbf{k}) = \frac{1}{M}\sum_{\mathbf{R}}\bar{\bar{\Phi}}(\mathbf{R})e^{-i\mathbf{k}\mathbf{R}}.
\end{equation}
for each wave vector $\mathbf{k}$ in the first Brillouin zone one
finds the eigenvalues $\omega_{\mathbf{k}s}$ and eigenvectors
$\mathbf{\epsilon}_{\mathbf{k}s}$ of different phonon modes
(longitudinal or transverse) labeled by the symbol $s$, $N$ being
the number of atoms. Introducing the canonical phonon coordinates
$\mathbf{U}_{\mathbf{R}}$ and $\mathbf{P}_{\mathbf{R}}$
\begin{eqnarray}
\mathbf{U}_{\mathbf{R}} = \frac{1}{\sqrt{MN}}\sum_{\mathbf{k},s}\mathcal{Q}_{\mathbf{k}s}\mathbf{\epsilon}_{\mathbf{k}s}e^{i\mathbf{k}\mathbf{R}} \label{eq:cantr1} \\
\mathbf{P}_{\mathbf{R}} =
\frac{1}{\sqrt{MN}}\sum_{\mathbf{k},s}\mathcal{P}_{\mathbf{k}s}\mathbf{\epsilon}_{\mathbf{k}s}e^{i\mathbf{k}\mathbf{R}}.
\label{eq:cantr2}
\end{eqnarray}
allows a separation of the original Hamiltonian of the crystal
into the Hamiltonians of $3N$ independent harmonic oscillators.

The thermodynamic average of the operators
$\mathcal{Q}^{\dagger}_{\mathbf{k}s}\mathcal{Q}_{\mathbf{k}s}$
determines the mean-square atomic displacements and is given by
\begin{equation}
\langle \mathcal{Q}^{\dagger}_{\mathbf{k}s}\mathcal{Q}_{\mathbf{k}s} \rangle =
\frac{\hbar}{\omega_{\mathbf{k}s}}\Big [ \frac{1}{2} + n \Big
(\frac{\hbar\omega_{\mathbf{k}s}}{k_{B}T} \Big ) \Big ],
\label{eq:thav}
\end{equation}
where $n(x) = 1/( e^{x}-1)$ is the Planck function. In the
classical limit, i.e for sufficiently high temperatures, the
operators $(1/\sqrt{M})\mathcal{Q}_{\mathbf{k}s}$ are
replaced by real numbers,
\begin{equation}\label{eq:newq}
\mathcal{A}_{\mathbf{k}s} \equiv \pm
\sqrt{\frac{\langle \mathcal{Q}^{\dagger}_{\mathbf{k}s}\mathcal{Q}_{\mathbf{k}s} \rangle}{M}}.
\end{equation}

Calculating the gradient of the potential energy in Eqn. \ref{eq:harmH1} with respect to the atomic 
displacements gives
the restoring force
\begin{equation}\label{eq:force}
\mathbf{F}_{\mathbf{R}}= - \sum_{R'}\bar{\bar{\Phi}}(\mathbf{R}-\mathbf{R}')\mathbf{U}_{\mathbf{R}'}.
\end{equation}
Fourier transforming Eqn. \ref{eq:force} and substituting $\mathbf{U}_{\mathbf{R}}$ with the expression in Eqn. 3
gives
\begin{equation}\label{eq:fforce}
\mathbf{F}_{\mathbf{k}} = -\sum_{s}M\omega_{\mathbf{k}s}^{2}\mathcal{A}_{\mathbf{k}s}\mathbf{\epsilon}_{\mathbf{k}s}.
\end{equation}
Finally, using the orthogonality of the eigenvectors $\epsilon_{{\bf k}s}$ the phonon frequencies can be expressed as
\begin{equation}\label{eq:omegeqh}
\omega_{\mathbf{k}s} = \Big [ -\frac{1}{M}\frac{ \mathbf{\epsilon}_{\mathbf{k}s}\mathbf{F}_{\mathbf{k}}}{\mathcal{A}_{\mathbf{k}s}} \Big ]^{1/2}.
\end{equation}

The equations discussed so far can be solved for dynamically stable materials, where each atom is located in a minimum of the function $U_{\bf R}$. 
It is important to note that this does not have to correspond to a global total energy minimum of the lattice, a local minimum suffices.
 For dynamically unstable materials $U_{\bf R}$ does not have a minimum at the lattice sites of the crystal structure. 
In this situation the equations discussed so far can not be used straight forwardly since they result in imaginary phonon frequencies.
 This represents a situation where the lattice under consideration spontaneously shifts atomic planes and/or atomic positions so that 
a new crystal structure lowers the total energy. We demonstrate the problem at hand by comparing in Fig.\ref{fig:firstp} the calculated (zero temperature) phonon 
spectra of the bcc phase of the group IVB elements with experimental data obtained at elevated temperatures. At these temperatures the group IVB elements 
are observed to be stable in the bcc crystal structure, and the measured phonon frequencies are naturally positive for all lattice vectors.
Fig.1 shows that calculations using a static bcc lattice result in a dynamically unstable situation with imaginary phonon frequencies. It should be noted that the failure
 describing the bcc phase of the group IVB elements using harmonic lattice theory (Fig.1, right column) { is not caused by any obvious error in the energy functional used, and are likely not to be improved even if an exact functional for a static lattice were found. 
\begin{figure*}[tbp]
\begin{center}
\includegraphics*[angle=0,scale=0.45]{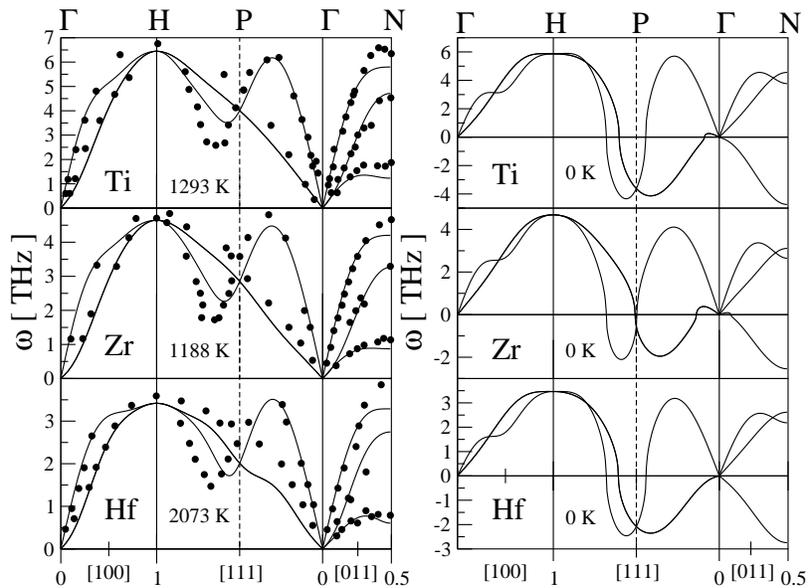}
\caption{ The phonon dispersions of the group IVB metals. The solid lines are the first principles self consistent 
phonon calculations. In the left column the finite temperature calculations, and in the right 
column the T = 0 K calculations. The filled circles are the experimental data of Ref. \cite{HEIMING,PETRY,TRAMENAU}.}
\label{fig:firstp}
\end{center}
\end{figure*}

In order to properly describe the high temperature phase of the group IVB elements on must include
the interaction
between phonons~\cite{ktsn}. 
As a result, phonon
frequencies turn out to be temperature dependent which we explore numerically in this study. However, we neglect
the phonon damping due to decay processes of
phonons (see, e.g., Ref.~\cite{CaSr} and Refs. therein), another anharmonic effect.
In the present  calculations thermal expansion effects have not
been taken into account, all calculations have been performed
at constant volume. Furthermore, the thermal excitations of the electronic subsystem  has 
not been considered in the present calculations of the phonon frequencies.

The method used to calculate temperature dependent phonons presented in this paper 
considers a supercell containing a number of atoms which are allowed to deviate from the lattice positions 
stipulated by the crystal structure. The deviations are calculated as a function of temperature, 
by solving equations  3 to 9 self consistently.
The deviation of the atomic positions away from the ideal lattice points provides an extra entropy to the system 
and the stabilization of the bcc structure for the group IVB elements as a function of increasing temperature may 
as we will see below be found. 
\begin{figure}[tbp]
\begin{center}
\includegraphics*[angle=0,scale=0.35]{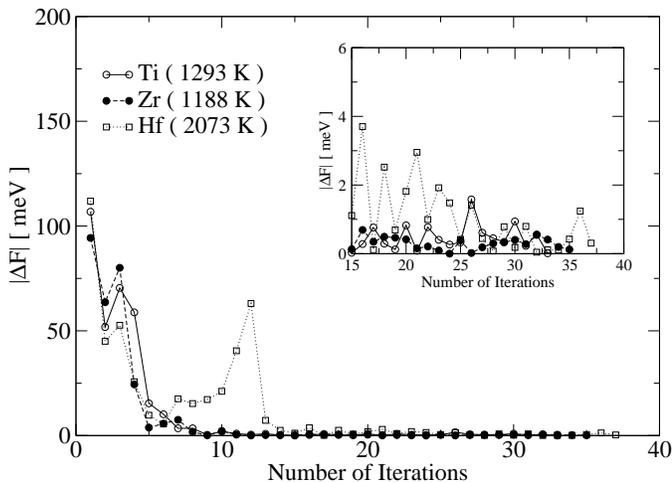}
\caption{The change in free energy between two consecutive iterations, here plotted as a function of the number of 
iterations. The inset in the figure shows the same plots but  at a smaller energy scale. }
\label{fig:conv}
\end{center}
\end{figure}

As regards the calculational details of the force calculation we used the VASP package \cite{VASP}, within the generalized
gradient approximation (GGA).
The PAW potentials used required
energy cutoffs of  197 eV for Ti, 175 eV for Zr,
and 243 eV for Hf. The k-point mesh was a 6x6x6 Monkhorst-Pack grid, 
and the supercell used was obtained by increasing the bcc primitive cell 4
times along the 3 primitive lattice vectors.

In practice our calculations are done by first calculating a starting guess for the phonon dispersions by means
of a standard supercell calculation, see e.g Ref. \onlinecite{PETROS}.
The phonon frequencies corresponding to k-vectors commensurate with the supercell
are then used to calculate the atomic displacements through Eqns. 3-6. Here it should be noted 
that the signs of the amplitudes $\mathcal{A}_{{\bf k}s}$ (see Eqn.\ref{eq:newq}), should be chosen randomly, with 
equal probabilities for + and -. This is an approximation to the procedure in which   $\mathcal{A}_{{\bf k}s}$ is sampled continuously to obtain the correct mean square deviations of the modes. This approximation however conserves correctly the property that the displacements in Eqn. 3 are {\bf R}-dependent. Furthermore, it should be noted that the eigenvectors $\mathbf{\epsilon}_{\mathbf{k}s}$ calculated in the initial calculation are not updated throughout the 
rest of the procedure. This however does not introduce any extra approximation, it merely guarantees that the longitudinal and transverse modes are fixed to the modes of the bcc lattice.
Starting from the equilibrium
geometry used in the initial supercell calculation, the atoms
are displaced according to Eq. 3, and the forces on these
displaced atoms are calculated. From the Fourier transform
of the atomic forces a new set of frequencies are calculated through Eqn (\ref{eq:omegeqh}).
To retain the correct symmetry of the calculated phonon dispersion the symmetries of
the different k-vectors are restored by
\begin{equation}\label{eq:symm}
\Omega_{\mathbf{k}s}^{2}= \frac{1}{m_{\mathbf{k}}}\sum_{\mathcal{S}\in \mathcal{S}(\mathbf{k})}\omega_{\mathcal{S}^{-1}\mathbf{k}s}^{2},
\end{equation}
where $\mathcal{S}(\mathbf{k})$ is the symmetry group of the wave
vector $\mathbf{k}$, and $m_{k}$ the number of elements of the
group. From the different iterations frequency distributions of the modes are
obtained, and a new set of frequencies are supplied by the mean frequencies of these distributions,
\begin{equation}\label{eq:mean}
\omega_{\mathbf{k}s}^{2}(N)= \frac{1}{N}\sum_{i=1}^{N}\Omega_{\mathbf{k}s}^{2}(i),
\end{equation}
where  $\Omega_{\mathbf{k}s}(i)$, $i=1,...,N$ are the symmetry
restored frequencies from all iterations. The new set of
frequencies calculated in (\ref{eq:mean}) determine a new set of
displacements used to calculate a new set of forces. 
Philosophically our approach is similar to Born's self consistent phonon theory, with the main difference being that we consider 
a direct force calculation from a super cell with Hellman-Feynman forces calculated from density functional  theory.



Figure \ref{fig:firstp} shows
the calculated phonon dispersions together with the experimental data of  Ref. \cite{HEIMING,PETRY,TRAMENAU}
for the bcc phase of the group IVB metals at temperatures 1293 K, 1188 K, and 2073 K for Ti, Zr, and Hf, respectively.
The finite temperature calculations predict the stability of the bcc phase of all
group IVB metals by promoting the frequencies of the phonons along the $\Gamma$ to $N$ symmetry line and around the $P$ symmetry point
from imaginary to real.
The finite
temperature calculations of phonons result in an overall quantitative agreement with experimental values. Smaller deviations are
observed around the P and H point of the Brillouin-zone, most likely due to finite size effects of the supercell used in the calculations.

From the self consistent phonon spectrum
the free energy is approximated
from the density of states of the phonons $g(\omega)$ through the expression
\begin{equation}\label{eq:phonF}
F(T) = \int_{0}^{\infty}d\omega g(\omega)[\frac{\hbar\omega}{2}+k_{B}Tln(1-e^{-\hbar\omega/k_{B}T})],
\end{equation}
which has been shown by Cochran {\it et al.} \cite{COCH} to give an entropy correct to leading order in anharmonic 
perturbation theory.
Figure \ref{fig:conv} shows the convergence of
free energy for the three elements considered in this work.
In all calculations presented here the self consistent cycle was terminated when the difference in the approximate free energy of the lattice
between two consecutive iterations was less than 1 meV. Convergence in the free energy with such accuracy is very encouraging and opens up the possibility to investigate temperature induced phase stability for a very large set of materials, since the accuracy needed to e.g. resolve crystallographic energy differences is of the order of a few meV or more. This prediction has also been tested by using inter-atomic forces calculated 
with the embedded atom potentials of Ref. \onlinecite{PAS1,PAS2}. The free energy difference 
between the hcp and bcc structures where calculated as functions of temperature for Ti and Zr. Here the theoretically 
predicted hcp to bcc transition temperatures where within $\sim$400 K of the corresponding experimental temperatures.



In summary, a quantitative theory
successfully
explains the long lasting question concerning thermal, entropy driven stabilization of dynamically unstable materials.
Application to the group IVB elements reproduces
the measured phonon spectrum of these elements at elevated temperatures with good accuracy. We note that the presented method reproduces 
observed high temperature phonon spectra with good accuracy and that the method when 
used at low, but non-zero temperatures, results in imaginary frequencies for e.g. bcc Ti. This shows that at low temperatures this element is unstable in the bcc phase, in agreement with observations.
Other systems where one can expect success of this method are
the bcc phase
of f-electron materials as well as, the high pressure phase of Fe, and many of the
ferroelectrics. The approach has advantages over
traditional methods such as MD simulations in that complications
associated with metallic materials are avoided and, most
importantly, that a much smaller set of atoms are needed.

\bigskip



\end{document}